\documentclass[lettersize,journal]{IEEEtran}
\usepackage{amsmath,amsfonts}
\usepackage{algorithmic}
\usepackage[ruled,linesnumbered]{algorithm2e}
\usepackage{array}
\usepackage{booktabs}  
\usepackage{tabularx}  
	\usepackage[caption=false,font=normalsize,labelfont=sf,textfont=sf]{subfig}
	\usepackage{textcomp}
	\usepackage{stfloats}
	\usepackage{url}
	\usepackage{balance}
	\usepackage{verbatim}
	\usepackage{graphicx}
	\usepackage[colorlinks]{hyperref}
	\hypersetup{citecolor=blue}
	\usepackage{threeparttable}
	\usepackage{multirow}
	\usepackage{amssymb}
	\usepackage{cite}
\begin{document}
	
	\title{Fundamental Tradeoffs for ISAC Multiple \\Access in Finite-Blocklength Regime}

\author{Zhentian Zhang, Christos Masouros, Kai-Kit Wong, Jian Dang, Zaichen Zhang, \\Kaitao Meng, Farshad Rostami Ghadi, Mohammad Javad Ahmadi
\thanks{}
\thanks{Z.-T. Zhang is with the National Mobile Communications Research Laboratory, Southeast University, Nanjing, 210096, China. (e-mail: zhentianzhangzzt@gmail.com).}
\thanks{J. Dang is with the National Mobile Communications Research Labo-ratory, Frontiers Science Center for Mobile Information Communication and Security, Southeast University, Nanjing 211189, China, also with the Key Laboratory of Intelligent Support Technology for Complex Environments, Ministry of Education, Nanjing University of Information Science and Tech-nology, Nanjing 210044, China, and also with Purple Mountain Laboratories, Nanjing 211111, China (e-mail: dangjian@seu.edu.cn).}
\thanks{Z.-C. Zhang is with the National Mobile Communications Research Laboratory, Frontiers Science Center for Mobile Information Communication and Security, Southeast University, Nanjing, 210096, China and he is also with the Purple Mountain Laboratories, Nanjing 211111, China (e-mail: zczhang@seu.edu.cn).}
\thanks{C. Masouros, K.-K. Wong, F. Rostami Ghadi are with the Department of Electronic and Electrical Engineering, University College London, Torrington Place, WC1E 7JE, United Kingdom and K.-K. Wong is also with the Yonsei Frontier Lab., Yonsei University, 03722 Korea  (e-mails: \{c.masouros, kai-kit.wong, f.rostamighadi\} @ucl.ac.uk).}
\thanks{K. Meng is with the Department of Electrical and Electronic Engineering, University of Manchester, Manchester, UK (email: kaitao.meng@manchester.ac.uk).}
\thanks{M. J. Ahmadi is with the Chair of Information Theory and Machine Learning, Technische Universität Dresden, 01062 Dresden, German (E-mail: mohammad\_javad.ahmadi@tu-dresden.de).}}

	%
	 \pagestyle{empty}
	\maketitle
	  \thispagestyle{empty}
	\begin{abstract}
This paper investigates the fundamental communication--sensing tradeoffs of uplink dual-functional integrated sensing and communication (ISAC) multiple access under finite blocklength (FBL) constraints. Unlike conventional asymptotic analyses, we explicitly account for the limitations under FBL constraints imposed by short packets and low-latency transmission. By examining the unbiased channel state sensing estimator, we establish a geometric decomposition of the sensing error, indicating that it is jointly determined by the signal-to-noise ratio and the correlation structure of the information codebook. This insight reveals how cross-correlation among active users in the codebook geometry fundamentally constrains dual-functional ISAC performance. Consequently, we derive achievability and converse bounds that characterize the tradeoff between communication code rate and sensing accuracy in the FBL regime, with the converse further bounded by Shannon capacity. Moreover, by treating channel state sensing as a high-level sensing objective, a universal Cram\'er--Rao bound is derived to link channel estimation accuracy to practical sensing parameters. Examples of parameter sensing are also provided based on 3GPP standard. Numerical results validate the theoretical analysis and demonstrate the impact of blocklength, antenna dimensions, and sensing requirements.

	\end{abstract}
	\begin{IEEEkeywords}
	 Dual-functional integrated sensing and communication, multiple access, finite blocklength, tradeoff.
	\end{IEEEkeywords}
	\section{Introduction}
Integrated Sensing and Communication (ISAC) \cite{ISAC1} transforms network design by fusing radar and communication via advanced signal processing \cite{ISAC3}. Central to this convergence is ISAC multiple access (MA), which manages mutual interference between functions \cite{ISAC_Mulitple_Access}. MA schemes are generally categorized into orthogonal (OMA) and non-orthogonal (NOMA) approaches. While OMA mitigates interference through explicit resource separation \cite{TMDA1}, it lacks the spectral efficiency and scalability of NOMA \cite{NOMA-ISAC}. Crucially, existing literature \cite{ISAC-fundamentals} predominantly assumes asymptotic resource allocation, neglecting practical finite blocklength (FBL) constraints \cite{Polyanskiy}. Since FBL coding prevents error-free communication and inherently degrades sensing precision in low-latency scenarios, analyzing the performance gap between asymptotic and FBL regimes is vital for realistic ISAC system design \cite{FBL}.

Motivated by the practical FBL constraints, recent research has begun to characterize fundamental sensing-communication tradeoffs in the FBL regime. Studies such as \cite{FBL-ISAC2, FBL-ISAC1, FBL-ISAC3} derive second-order bounds for point-to-point and single-input single-output (SISO) systems, revealing how sensing requirements constrain the information codebook structure. However, these investigations are largely restricted to single-user or simple orthogonal setups. Consequently, the analysis of ISAC MA techniques, specifically how to manage multi-user interference under strict FBL constraints, remains a critical open challenge to be explored.

This work aims to characterize the fundamental tradeoffs of integrated sensing and communication (ISAC) multiple access under finite blocklength (FBL) constraints. The main contributions are summarized as follows:
\begin{itemize}
	\item We first uncover the geometric relationship between channel-state sensing and communication inherent in the structure of unbiased estimators. Our analysis shows that, in ISAC multiple access under the FBL regime, the correlation among users' codewords plays a critical role in determining dual-functional performance. This geometric perspective on the information codebook enables a systematic characterization of communication--sensing tradeoffs.
	
	\item We then establish achievability and converse bounds that characterize the tradeoff between code rate and sensing error for the dual-functional ISAC multiple access in the FBL regime. These bounds reveal how sensing requirements impose additional constraints on random coding codebooks, thereby fundamentally shaping the interplay between communication rate and sensing accuracy.
	
	\item Finally, to validate the effectiveness of treating channel estimation as a high-level sensing objective, we derive a universal Cramér--Rao bound (CRB) for generic sensing parameters. Based on the 3GPP standard, practical sensing parameters, including angle of arrival (AoA), range, and velocity, are further examined as illustrative examples of the derived lower bounds.
\end{itemize}

The remainder of this paper is organized as follows. Sec.~\ref{sec:tradeoff_analysis} introduces the dual-functional ISAC system model and presents the fundamental communication--sensing tradeoff bounds. Sec.~\ref{sec:param_mapping} derives a universal CRB for practical sensing parameter estimation and details the corresponding Jacobian derivations for parameters such as AoA, range, and velocity. Numerical results are provided in Sec.~\ref{sec.VI}, and concluding remarks are drawn in Sec.~\ref{sec.VII}.

{\em Notations:} Capital letters denote matrices (e.g., $A$), with $A_{i,:}$ and $A_{:,j}$ representing the $i$-th row and $j$-th column, respectively. Lowercase letters denote real or complex scalars, while $\mathbb{R}$ and $\mathbb{C}$ denote the real and complex domains. Calligraphic letters denote sets or functions (e.g., $\mathcal{A}$). Common operators include the expectation $\mathbb{E}[\cdot]$, the $\ell_2$ norm $\|\cdot\|_2$, the transpose $(\cdot)^T$, and the Hermitian transpose $(\cdot)^H$. The symbol $\mathbf{0}$ denotes an all-zero matrix, $\mathrm{diag}(X)$ a diagonal matrix with entries $X$, and $\mathfrak{Re}\{\cdot\}$ the real part of a complex scalar.

\section{Fundamental Tradeoff Analysis: \\Achievability and Converse}\label{sec:tradeoff_analysis}
\subsection{System Configurations}
We consider an uplink transmission scenario where a dual-functional receiver equipped with $m$ antennas serves $k$ active single-antenna users. Each user transmits $b$ information bits over $n$ channel uses (e.g., time slots or frequency resource blocks).  Without loss of generality, we assume the existence of a transmitter design (including coding, waveform, and transmission structure) that maps $b$ bits into a common codebook $A \in \mathbb{C}^{2^b \times n}$, whose entries are independently distributed as $\mathcal{CN}(0,\bar{p})$, where $\bar{p}$ denotes the average transmit power per channel use. Among the $2^b$ codewords, only $k$ rows are active, forming the transmitted signal matrix $X \in \mathbb{C}^{k \times n}$.  Let $H \in \mathbb{C}^{m \times k}$ denote the channel matrix corresponding to the active users, whose entries have variance $\sigma_H^2$. Neglecting asynchronous impairments, the received signal at the receiver is given by
\begin{equation}\label{eq:linear_MIMO}
	Y = H X + N,
\end{equation}
where $H$ is treated as a deterministic matrix embedding sensing-related parameters (e.g., distance, angle, and velocity) and is estimated after decoding $X$ from $Y$. The noise matrix $N \in \mathbb{C}^{m \times n}$ is modeled as additive white Gaussian noise (AWGN) with independent entries distributed as $\mathcal{CN}(0,\sigma_n^2)$. In practice, different users experience distinct fading conditions, resulting in unequal received powers. However, since this work focuses on accommodating a large number of users in a dual-functional system, we adopt a quasi-static multiple-access model with identical power constraints across users. The energy-per-bit metric is defined as $\frac{E_b}{N_0} = \frac{n\bar{p}}{b\sigma_n^2}$, where $\bar{p}$ is the transmit power per channel use.

\subsection{Bridging Sensing and Communication}
Here, we derive the fundamental tradeoff between the communication rate $r=\frac{b}{n}$ and the sensing accuracy, i.e., normalized mean squared error (NMSE). Specifically, we map the estimation performance to the geometric properties of the transmitted codewords in the high-dimensional Euclidean space. This mapping then reveals how the size increment of the codebook, controlled by $b$, inevitably degrades the condition number of the sensing matrix, thereby establishing the theoretical limits, i.e. the achievability and the converse tradeoff bounds.

The estimation error from the unbiased least-squares (LS) estimator equals to $\Delta_{LS} = \widehat{H} - H = N X^H (XX^H)^{-1}$. Therefore, the NMSE can be equivalently calculated as:
\begin{equation}
	\mathrm{NMSE} = \mathbb{E} \left[ \frac{\| \Delta_{LS} \|_F^2}{\| H \|_F^2} \right] = \frac{1}{m k \sigma_H^2} \mathbb{E} \left[ \mathrm{tr}\left( \Delta_{LS} \Delta_{LS}^H \right) \right].
\end{equation}
Substituting $\Delta_{LS} = N X^H G^{-1}$, where $G = XX^H$ is the Gram matrix, and using the identities of $\mathrm{tr}(AB) = \mathrm{tr}(BA)$ and the AWGN independence of $\mathbb{E}[N^H N] = m \sigma_n^2 I_n$:
\begin{equation}\label{eq:NMSE_geometry_decomp}
	\mathrm{NMSE} = \underbrace{\left( \frac{\sigma_n^2}{n\bar{p}\sigma_H^2} \right)}_{e_{\min}} \cdot \underbrace{\left( \frac{n\bar{p}}{k} \mathrm{tr}(G^{-1}) \right)}_{G_{\eta}}.
\end{equation}
Some important factors can be observed from \eqref{eq:NMSE_geometry_decomp}:
\begin{itemize}
	\item The term $e_{\min}$ depends only on the signal-to-noise ratio (SNR) at single antenna and blocklength $n$. It represents the lowest sensing error achievable \emph{only if} the signal matrix $X$ is perfectly orthogonal, i.e., $G = n\bar{p}I$.
	\item The term $G_{\eta}$ implies the penalty caused by the non-orthogonality of the active users' codewords. Geometrically, if the $k$ codewords from span of $\mathbb{C}^n$ have small angles between them, i.e., highly correlated, the Gram matrix $G$ becomes ill-conditioned. Consequently, its eigenvalues $\ell_i$ drop, and $\mathrm{tr}(G^{-1}) = \sum\ell_i^{-1}$ surges.
\end{itemize}

Thereby, the sensing performance is strictly dictated by the geometry of the codewords in the information codebook. Before elaboration on the proposed tradeoff bounds,
based on \cite[Theorem~4]{Theorem3}, one could feasibly calculate and approximate the maximal codeword correlation as:
\begin{subequations}\label{eq:rho}
	\begin{align}
		\rho_{\max} &= \sqrt{\frac{\ln t}{n}}+\gamma\frac{1}{2\sqrt{n\ln t}},~t=\frac{2^b\left(2^b-1\right)}{2},\label{eq:rho_closed}\\
		&\approx \underbrace{\sqrt{\frac{2b\ln 2}{n}}}_{\mathcal{F}(b)}, \mathrm{if}~2^b\gg 1,\label{eq:rho_link}
	\end{align}
\end{subequations}
which defines a monotonic bijection function $\rho_{\max} = \mathcal{F}(b)$ in \eqref{eq:rho_link}. In the following analyses, we invert this function to determine the maximum supportable bits $b$ for a given correlation constraint.

\subsection{Derivation of the Tradeoff Achievability Bound}
The achievability bound seeks to determine the maximum communication rate $\frac{b}{n}$ that can be supported while guaranteeing the sensing requirement $\mathrm{NMSE} \le e_{th}$ is met, where $e_{th}$ is the sensing error threshold, i.e., if $\mathrm{NMSE}\le e_{th}$, what is the achievable code rate $\frac{b}{n}$ to the very least. This requires a {\em worst-case} analysis of the geometry factor $G_{\eta}$ in \eqref{eq:NMSE_geometry_decomp}.

We analyze the worst-case geometric configuration for a given maximum correlation $\rho_{\max}$, where the sensing error peaks as the Gram matrix $G$ approaches singularity. To quantify this, we invoke the Gershgorin Circle Theorem \cite[Chapter~6]{GCT}, which bounds the eigenvalues based on the matrix entries. Specifically, the theorem implies that the minimum eigenvalue $\ell_{\min}$ is lower-bounded by the diagonal entry, e.g., signal energy $n\bar{p}$ in our case, minus the sum of the absolute off-diagonal entries, e.g., $|\rho_{ij}|$ absolute inter-correlation value in our case. By imposing the worst-case assumption where all pairwise correlations reach $\rho_{\max}$, we obtain:
\begin{equation}\label{eq.worst_condition}
	\ell_{\min} \ge n\bar{p} \left(1 - \sum_{j \neq i} |\rho_{ij}|\right) \ge n\bar{p} \big(1 - (k-1)\rho_{\max}\big).
\end{equation}

Furthermore, substituting this worst-case eigenvalue of \eqref{eq.worst_condition} into the NMSE expression \eqref{eq:NMSE_geometry_decomp} by replacing the trace sum $\sum\ell_i^{-1}$ with $k \ell_{\min}^{-1}$:
\begin{equation}
	\begin{aligned}
			\mathrm{NMSE}_{\max} \le&~ e_{\min} \frac{n\bar{p}}{k} \frac{k}{n\bar{p}(1 - (k-1)\rho_{\max})} \\
			&= \frac{e_{\min}}{1 - (k-1)\rho_{\max}}.
	\end{aligned}
\end{equation}
Setting $\mathrm{NMSE}_{\max} \le e_{th}$, the permissible correlation $\widehat{\rho}_{\mathrm{achi}}$ is:
\begin{equation}\label{eq:perm_cor_achi}
	\begin{aligned}
		&\frac{e_{\min}}{1 - (k-1)\rho_{\max}} \le e_{th}\\ \implies&~ \rho_{\max} \le \underbrace{\frac{1}{k-1} \left( 1 - \frac{e_{\min}}{e_{th}} \right)}_{\text{permissible correlation}~\widehat{\rho}_{\mathrm{achi}}}.
	\end{aligned}
\end{equation}
By adopting the relationship between the number of bits $b$ and permissible correlation in \eqref{eq:rho}, i.e., $b=\mathcal{F}^{-1}\left(\widehat{\rho}_{\mathrm{achi}}\right)$,  we obtain the achievability boundary as $\mathcal{B}_{\mathrm{achi}}\left(e_{th},\frac{\mathcal{F}^{-1}\left(\widehat{\rho}_{\mathrm{achi}}\right)}{n}\right)$, where $\mathcal{B}_{\mathrm{achi}}$ indicates a function with input $e_{th}$ and output $\frac{\mathcal{F}^{-1}\left(\widehat{\rho}_{\mathrm{achi}}\right)}{n}$. Any region below this curve ensures that even in the worst geometric alignment of users, the sensing precision is maintained.

\subsection{Derivation of the Tradeoff Converse Bound}

The converse bound reveals the fundamental limits of sensing performance for a random codebook of size $2^b$, leveraging the statistical properties of random coding ensembles. This provides an upper bound on $b$ that cannot be exceeded by standard random coding constructions with sensing constraint. Unlike the worst-case scenario, typical random codeword vectors in high-dimensional space are not aligned to produce the worst-case interference. Instead, the off-diagonal elements of the Gram matrix $G = n\bar{p}(I + \Delta)$ behave like incoherent noise where $\Delta$ represents the normalized off-diagonal correlation matrix with zero diagonal. We utilize the Neumann series expansion, i.e., $\left(I-T\right)^{-1}=\sum_{k=0}^{\infty}T^k$, for the matrix inverse of $(I+\Delta)^{-1} \approx I - \Delta + \Delta^2$. The trace of the inverse can be simplified as follows:

\begin{equation}
	\begin{aligned}
		\mathrm{tr}(G^{-1}) &= \frac{1}{n\bar{p}} \mathrm{tr}\left( (I_k + \Delta)^{-1} \right) \\
		&\approx \frac{1}{n\bar{p}} \left( \underbrace{\mathrm{tr}(I_k)}_{\text{Dimension } k} - \underbrace{\mathrm{tr}(\Delta)}_{\text{Zero}} + \underbrace{\mathrm{tr}(\Delta^2)}_{\text{Correlation Energy}} \right),
	\end{aligned}
\end{equation}
where $\mathrm{tr}(\Delta) = 0$ because the diagonal elements of $\Delta$ are zero by definition and self-correlations are captured in $I_k$. Furthermore, since $G$ is Hermitian, $\Delta$ is also Hermitian, i.e., $\Delta_{i,j} = \Delta_{j,i}^*$. Thereby, the trace of the square can be linked to the Frobenius norm:
\begin{equation}
	\begin{aligned}
			\mathrm{tr}(\Delta^2) &= \sum_{i=1}^k (\Delta^2)_{i,i} = \sum_{i=1}^k \sum_{j=1}^k \Delta_{i,j} \Delta_{j,i} \\
			&= \sum_{i,j} |\Delta_{ij}|^2 = \|\Delta\|_F^2.
	\end{aligned}
\end{equation}
Consequently, we can obtain the approximation dependent on the total correlation energy:
\begin{equation}
	\mathrm{tr}(G^{-1}) \approx \frac{1}{n\bar{p}} \left( k + \|\Delta\|_F^2 \right),
\end{equation}
where the term $\|\Delta\|_F^2$ represents the total energy of the cross-correlations. 

For a codebook with maximum correlation $\rho_{\max}$, the sum of squared off-diagonal elements is tightly concentrated around $k(k-1)\rho_{\max}^2$. Substituting this back into the geometry factor $G_{\eta}$:
\begin{equation}\label{eq:panlty_conv}
	G_{\eta} \approx 1 + \frac{1}{k} \|\Delta\|_F^2 \le 1 + (k-1)\rho_{\max}^2.
\end{equation}
Notably, the penalty in \eqref{eq.worst_condition} scales linearly with $\rho_{\max}$ in the worst case for achievability bound, but quadratically with $\rho_{\max}$ in the typical case of converse bound in \eqref{eq:panlty_conv}. Clearly, these outcomes are predicted to pose advantages for the converse bound since the correlation constant is less than 1.

Substituting \eqref{eq:panlty_conv} into the NMSE expression in \eqref{eq:NMSE_geometry_decomp} and setting the sensing lower bound to $e_{th}$:
\begin{equation}\label{eq.converse_cor}
	\begin{aligned}
	&\mathrm{NMSE}_{\mathrm{converse}} \le e_{\min} \big(1 + (k-1)\rho_{\max}^2\big) \le e_{th},\\
	&\implies \rho_{\max} \le \underbrace{\sqrt{\frac{1}{k-1} \left( \frac{e_{th}}{e_{\min}} - 1 \right)}}_{\widehat{\rho}_{\mathrm{conv}}},
\end{aligned}
\end{equation}
where $e_{\min}$ is a constant in \eqref{eq:NMSE_geometry_decomp}.
Finally, converting the $\rho_{\max}$ in \eqref{eq.converse_cor} to bits $b$ gives the converse bound by $\mathcal{B}_{\mathrm{conv}}\left(e_{th},\frac{\mathcal{F}^{-1}\left(\widehat{\rho}_{\mathrm{conv}}\right)}{n}\right)$. The value $\frac{\mathcal{F}^{-1}\left(\widehat{\rho}_{\mathrm{conv}}\right)}{n}$ defines the theoretical upper bound on the code rate. Consequently, with appropriate optimization, any code rate below this limit allows the system to satisfy the sensing error threshold $e_{th}$.

Moreover, an uplink multi-antenna system achieves its fundamental capacity limit in a rich scattering environment without sensing constraints. To upper-bound the performance, we employ the Shannon capacity as the ceiling for the converse bound. Specifically, we consider a Rayleigh fading channel model where the entries of $H \in \mathbb{C}^{m \times k}$ are i.i.d. with zero mean and variance $\sigma_H^2$. Assume that each user transmits with an individual power budget, denoted by the per-user signal-to-noise ratio $\mathrm{SNR}$. The ergodic sum capacity \cite[Chapter~8]{Funda_wireless} is bounded by:
\begin{equation}\label{eq.shannon}
	\frac{\widehat{b}_{\mathrm{tot}}}{n} \le \mathbb{E} \left[ \log_2 \det \left( I_m + \mathrm{SNR} HH^H \right) \right],
\end{equation}
where $\widehat{b}_{\mathrm{tot}}$ denotes the total information bits from all $k$ users. Consequently, the maximum single-user code rate for the converse bound is given by $\frac{\widehat{b}_{\mathrm{tot}}}{nk}$.

By considering the concavity of $\log\det(\cdot)$ \cite{Concavity} and applying Jensen's inequality, i.e., $\mathbb{E}[\log \det(\cdot)] \le \log \det(\mathbb{E}[\cdot])$, we simplify \eqref{eq.shannon} based on the relationship between the number of receive antennas $m$ and users $k$:
\begin{itemize}
	\item $m \ge k$: The system performance is limited by the number of independent spatial streams (i.e., users $k$). Using the identity $\det(I_m + AB) = \det(I_k + BA)$, we rewrite the bound as:
	\begin{equation}
		\frac{\widehat{b}_{\mathrm{tot}}}{n} \le \mathbb{E}\left[ \log_2 \det \left( I_k + \mathrm{SNR} H^H H \right) \right].
	\end{equation}
	Applying Jensen's inequality requires computing $\mathbb{E} [H^H H]$. The diagonal elements represent the sum of channel gains from a specific user $i$ to all $m$ receive antennas (capturing the receive array gain):
	\begin{equation}
		\mathbb{E} [(H^H H)_{i,i}] = \mathbb{E} \left[ \sum_{l=1}^m |h_{l,i}|^2 \right] = \sum_{l=1}^m \sigma_H^2 = m\sigma_H^2.
	\end{equation}
	Since the off-diagonal elements have zero mean, it follows that $\mathbb{E} [H^H H] = m\sigma_H^2 I_k$. Thus, \eqref{eq.shannon} becomes:
	\begin{equation} \label{eq:cap_bound_case1}
		\frac{\widehat{b}_{\mathrm{tot}}}{n} \le k \log_2 \left( 1 + m \sigma_H^2 \mathrm{SNR} \right), \quad \text{if } m \ge k.
	\end{equation}
	
	\item $m < k$: The system is limited by the dimensions of the receiver $m$. Jensen's inequality can be directly applied to \eqref{eq.shannon}:
	\begin{equation}
		\frac{\widehat{b}_{\mathrm{tot}}}{n} \le \log_2 \det \left( I_m + \mathrm{SNR} \mathbb{E} [HH^H] \right).
	\end{equation}
	Here, the diagonal elements represent the aggregate power from all $k$ users arriving at a single receive antenna $j$:
	\begin{equation}
		\mathbb{E} [(HH^H)_{j,j}] = \mathbb{E} \left[ \sum_{l=1}^k |h_{j,l}|^2 \right] = \sum_{l=1}^k \sigma_H^2 = k\sigma_H^2.
	\end{equation}
	With $\mathbb{E}[HH^H] = k\sigma_H^2 I_m$, the expression simplifies to:
	\begin{equation} \label{eq:cap_bound_case2}
		\frac{\widehat{b}_{\mathrm{tot}}}{n} \le m \log_2 \left( 1 + k \sigma_H^2 \mathrm{SNR} \right), \quad \text{if } m < k.
	\end{equation}
\end{itemize}
Overall, the converse bound is summarized as:
\begin{equation}
	\min\left(\underbrace{\mathcal{B}_{\mathrm{conv}}\left(e_{\mathrm{th}},\frac{\mathcal{F}^{-1}\left(\widehat{\rho}_{\mathrm{conv}}\right)}{n}\right)}_{\text{around \eqref{eq.converse_cor}}}, \frac{\widehat{b}_{\mathrm{tot}}}{kn}\right).
\end{equation}

\section{From Channel Estimation \\to Parameter Estimation} \label{sec:param_mapping}

While the fundamental tradeoff analysis in Sec. \ref{sec:tradeoff_analysis} evaluates the reconstruction quality of the channel matrix $H$, practical ISAC applications prioritize extracting physical parameters embedded within $H$ (e.g., AoA, range, Doppler). In this section, we derive a universal lower bound for parameter estimation based on the channel estimate $\widehat{H}$ and analyze the impact of transmission blocklength $n$ and antenna number $m$.

\subsection{Universal Parameter Sensing Lower Bound}
We postulate the channel matrix $H$ is a differentiable function $H(P)$ of the sensing parameters $P = [\alpha_1, \dots, \alpha_k]^T \in \mathbb{R}^k$. Defining the mapping $\mathcal{V}(P) = \mathrm{vec}(H(P))$, the Jacobian matrix $J \in \mathbb{C}^{mk \times k}$ is given by $J = \partial \mathcal{V}(P) / \partial P^T$. 
Recalling the unbiased LS estimate $\widehat{H}$ with error $E_h$, we relate the parameter estimation error $E_p = \widehat{P} - P$ to vectorized-channel estimation error $E_h$ via a first-order Taylor expansion around the true parameter $P$:
\begin{equation}\label{eq.vec_error}
	\begin{aligned}
		\mathrm{vec}(\widehat{H}) &= \mathrm{vec}(H)+E_h=\mathcal{V}(P + E_p) \\
		&\approx \mathcal{V}(P) + \underbrace{\left. \frac{\partial \mathcal{V}(\zeta)}{\partial \zeta^T} \right|_{\zeta=P}}_{J} E_p.
	\end{aligned}
\end{equation}
Subtracting $\mathrm{vec}(H)$ from both sides yields the linearized relationship $E_h \approx J E_p$, valid in high SNR regimes where higher-order terms $\mathcal{O}(\|E_p\|_2^2)$ vanish. 
Under the linear Gaussian model, the parameter estimation MSE is lower-bounded by the CRB, $\mathrm{tr}(C_p) \ge \mathrm{tr}(F^{-1})$. The Fisher information matrix (FIM) is $F = 2 \mathfrak{Re}\{ J^H C_h^{-1} J \}$. Utilizing the property $\mathrm{vec}(N X^H G^{-1}) = ((G^{-1})^T (X^H X)^T (G^{-1})^* \otimes I_m) \mathrm{vec}(N)$ and the identity $(G^{-1})^T = (G^*)^{-1}$, the covariance of $E_h$ is derived as:
\begin{equation}
	C_h = \mathbb{E}[E_h E_h^H] = \sigma_n^2 \left( (G^*)^{-1} \otimes I_m \right).
\end{equation}
Substituting $C_h^{-1}$ into the FIM expression, the MSE bound is:
\begin{equation}\label{eq:MSE_param}
	\mathbb{E}\left[ \|\widehat{P} - P\|_2^2 \right] \ge \frac{\sigma_n^2}{2} \mathrm{tr}\left( \left( \mathfrak{Re}\left\{ J^H (G^* \otimes I_m) J \right\} \right)^{-1} \right).
\end{equation}
This bound highlights the interplay between sensing geometry ($J$) and communication signal geometry ($G$).

\textit{Impact of Receiving Antennas $m$:}
Increasing $m$ enhances the column norms of $J$ via coherent accumulation. As $J^H (G^* \otimes I_m) J$ scales linearly with $m$, the MSE reduces proportionally to $1/m$, allowing large arrays to compensate for power constraints.

\textit{Impact of Blocklength $n$:}
The blocklength affects performance via the Gram matrix $G$:
\begin{itemize}
	\item \textit{Case $k \ll n$:} Codewords are nearly orthogonal ($G \approx n\bar{p}I_k$), making $G^* \otimes I_m$ a scaled identity. Parameter estimation is decoupled across users, limited only by SNR and $m$.
	\item \textit{Case $k \to n \text{~or~} k \ge n$:} Increased user density introduces cross-correlations, rendering $G$ ill-conditioned. The reduced minimum eigenvalue of $G$ degrades the FIM, reflecting severe inter-user interference in the parameter domain.
\end{itemize}
\begin{figure}[t!]
	\centering
	\includegraphics[width=0.7\columnwidth]{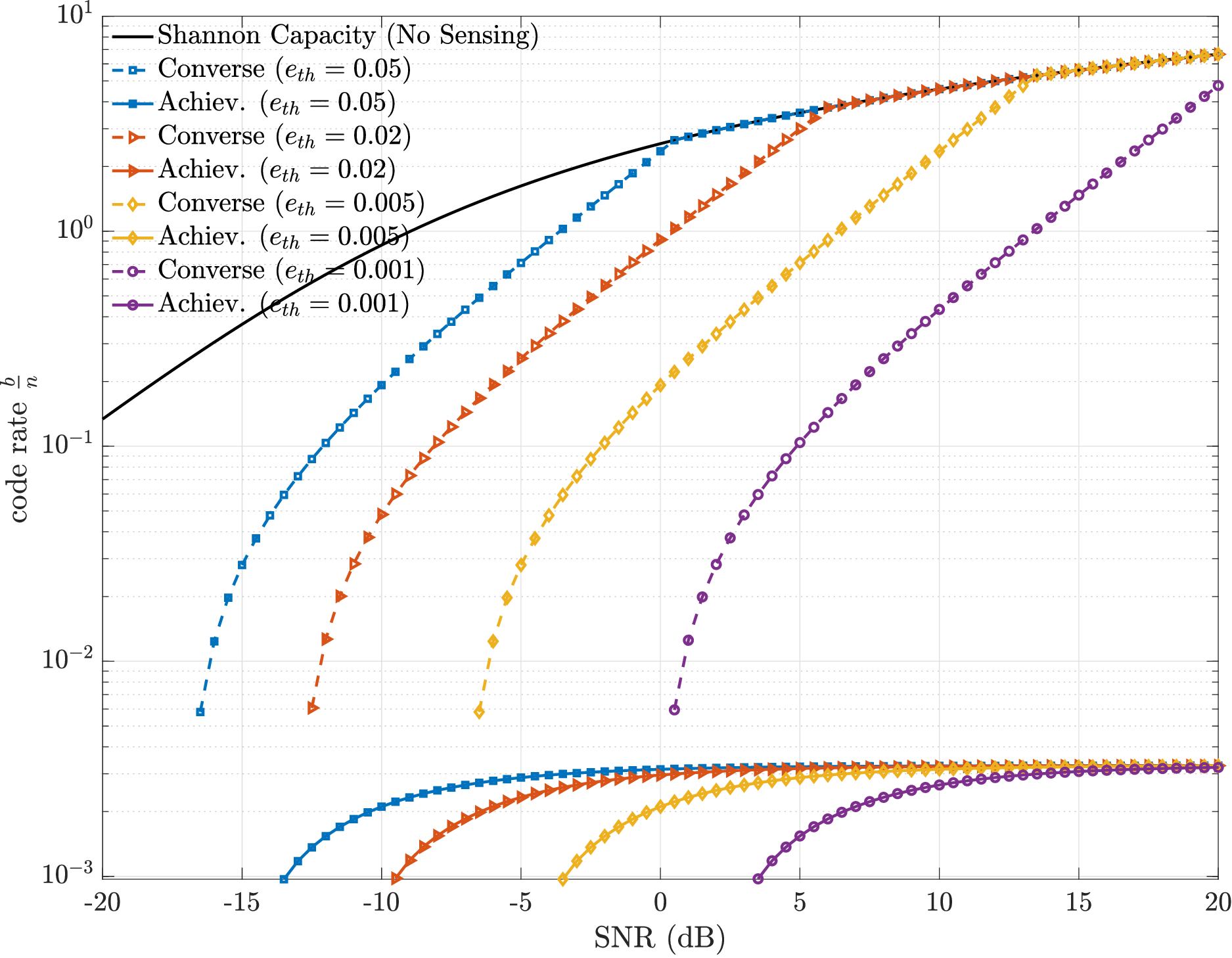}
	\caption{{\bf Tradeoffs:} Code rate achievability and converse bound versus SNR under fixed sensing error $e_{th}\in \{10^{-3},10^{-2},5\times 10^{-3},2\times 10^{-2},5\times 10^{-2}\}$, $k=16$ active users, $n=1000$ channel uses, $m=10$ receiving antenna.}
	\label{fig:coderate_SNR_fixed_sensing}
\end{figure}
\begin{figure*}[t!]
	\centering
	\subfloat[{\bf Achievability Bound:} $r$ vs $e_{th}$ vs $n$]
	{
		\includegraphics[width=0.8\columnwidth]{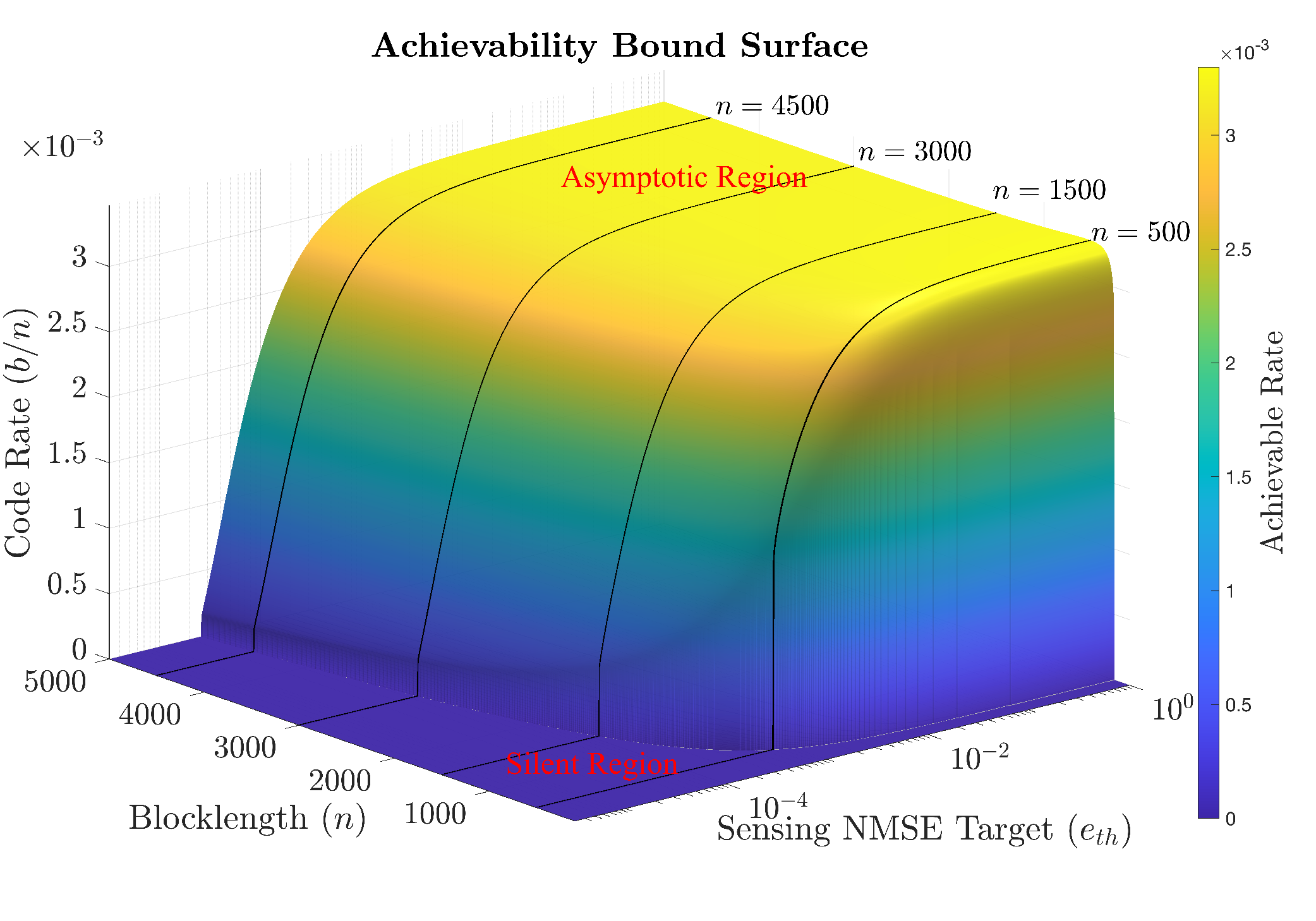}
		\label{fig:sub_achi}
	}
	\subfloat[{\bf Converse Bound:} $r$ vs $e_{th}$ vs $n$]{
		\includegraphics[width=0.8\columnwidth]{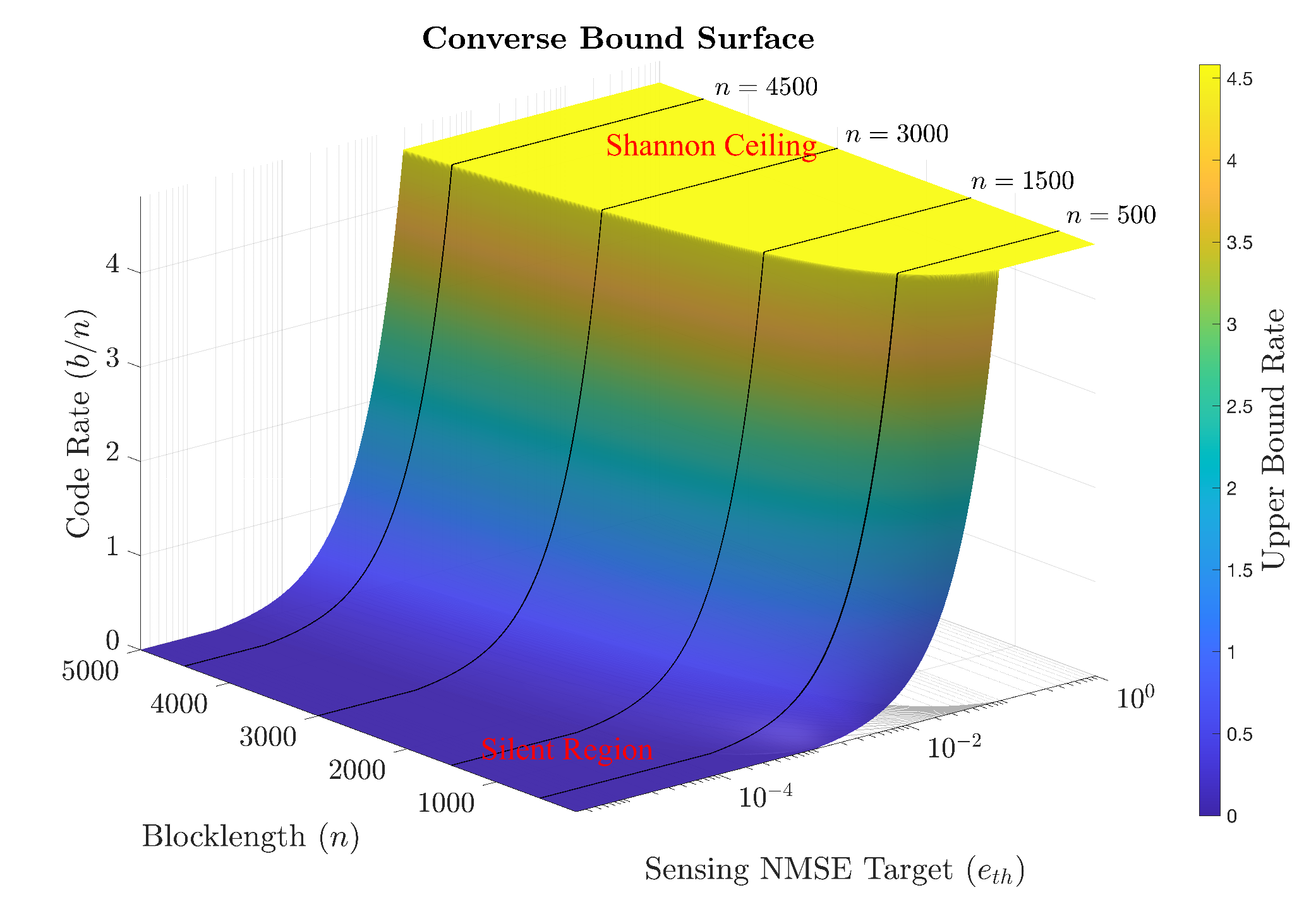}
		\label{fig:sub_conv}
	}
	\caption{Illustrations of how blocklength affects the communication--sensing tradeoff with $k=16$ active users, $m=10$ receiving antennas, and $\mathrm{SNR}=10\,\mathrm{dB}$.}
	\label{fig:achi_conv_wall}
\end{figure*}
\begin{figure*}[t!]
	\centering
	\includegraphics[width=1.6\columnwidth]{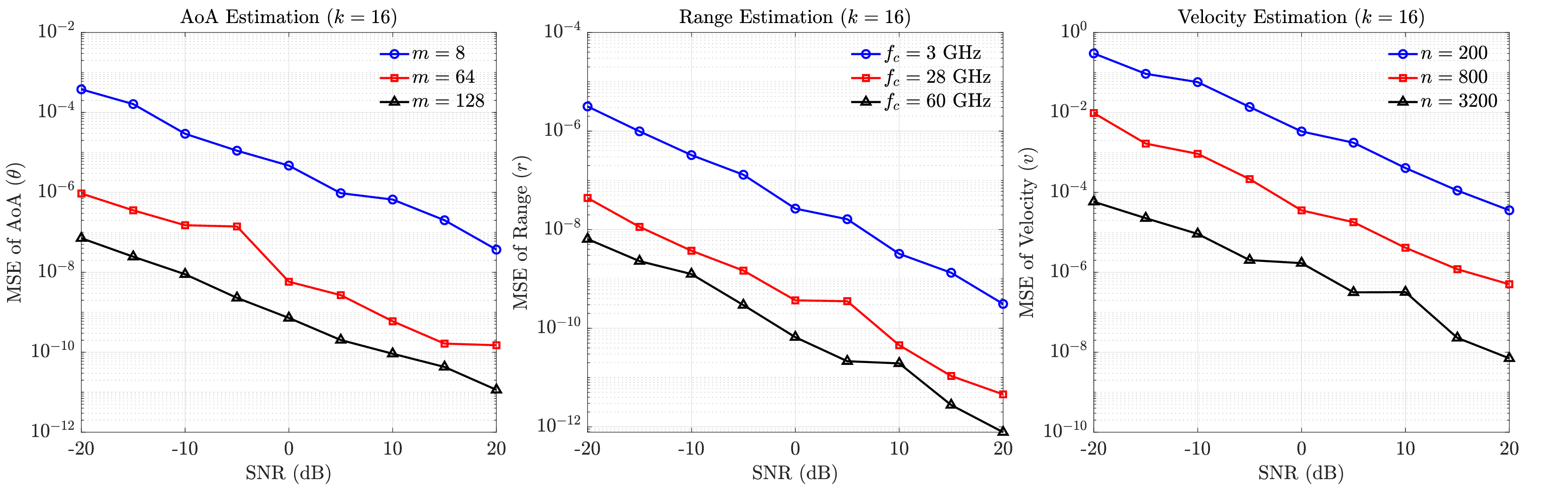}
	\caption{Illustrations on parameter estimation MSE lower bounds including AoA, range and velocity. Detailed parameter setups are listed in Tab.~\ref{tab:sim_params}.}
	\label{fig:practical_parameter_estimation}
\end{figure*}
\subsection{Practical Sensing Examples by 3GPP Standard}\label{sec.3GPP}

To instantiate the Jacobian derivation, we adopt the 3GPP TR 38.901 \cite{3gpp_38901} channel model. The line-of-sight (LoS) channel coefficient \cite{channel_modeling} with half-wavelength port placement for the $i$-th user at the $j$-th antenna is:
\begin{equation}\label{eq:3gpp_model}
	\begin{aligned}
		[H]_{j,i} = &\beta_i \underbrace{\exp\left(-j \frac{2\pi}{\lambda} d_a (j-1) \sin(\theta_i)\right)}_{\text{Spatial Term}}\times\\
		&\underbrace{\exp\left(-j \frac{4\pi f_c}{c} r_i\right)}_{\text{Range/Phase Term}}\times
		\underbrace{\exp\left(j \frac{4\pi f_c v_i}{c} t_{\text{obs}}\right)}_{\text{Doppler Term}}.
	\end{aligned}
\end{equation}
The Jacobian block $J_{:,i} = [\frac{\partial H_{:,i}}{\partial \theta_i}, \frac{\partial H_{:,i}}{\partial r_i}, \frac{\partial H_{:,i}}{\partial v_i}]$ is constructed by differentiating \eqref{eq:3gpp_model}. The partial derivatives with respect to the user state $P_i = [\theta_i, r_i, v_i]^T$ are:

\begin{enumerate}
	\item \textit{AoA $\theta_i$:} The derivative is $\frac{\partial [H]_{j,i}}{\partial \theta_i} = [H]_{j,i} \cdot (-j \pi (j-1) \cos(\theta_i))$. The sensitivity scales with the antenna index $j$, proving that larger apertures improve angular resolution.
	\item \textit{Range $r_i$:} The derivative is $\frac{\partial [H]_{j,i}}{\partial r_i} = [H]_{j,i} \cdot (-j \frac{4\pi}{\lambda})$. Sensitivity scales with $1/\lambda$, implying higher frequency bands provide superior phase resolution.
	\item \textit{Velocity $v_i$:} The derivative is $\frac{\partial [H]_{j,i}}{\partial v_i} = [H]_{j,i} \cdot (j \frac{4\pi f_c}{c} t_{\text{obs}})$. With $t_{\text{obs}} \approx n T_s$, velocity accuracy is directly proportional to the blocklength $n$ due to phase accumulation.
\end{enumerate}
Consequently, minimizing estimation error requires shaping the signal covariance $R_x$ to align the signal power with the subspace spanned by these sensitivity derivatives.

\begin{table}[htp]
	\centering
	\caption{Parameter Estimation Setups of Fig.~\ref{fig:practical_parameter_estimation}}
	\label{tab:sim_params}
	\renewcommand{\arraystretch}{0.8} 
	\begin{tabularx}{\columnwidth}{@{}X l@{}} 
		\toprule
		\textbf{Parameter} & \textbf{Value} \\ 
		\midrule
		
		\multicolumn{2}{@{}l}{\textit{\textbf{Default Settings}}} \\ 
		\midrule
		Speed of Light ($c$) & $3 \times 10^8$ m/s \\ 
		Noise Variance ($\sigma_n^2$) & 1 \\ 
		Channel Variance ($\sigma_H^2$) & 1 \\ 
		Symbol Duration ($T_s$) & $4~\mu$s \\ 
		Carrier Frequency ($f_c$) & 28 GHz \\ 
		Receive Antennas ($m$) & 10 \\ 
		Channel Uses ($n$) & 1000 \\ 
		Active Users ($k$) & 16 \\ 
		\midrule
		
		\multicolumn{2}{@{}l}{\textit{\textbf{Variations for Parameter Estimation}}} \\ 
		\midrule
		AoA MSE Analysis & $m \in \{8, 64, 128\}$ \\ 
		Range MSE Analysis & $f_c \in \{3, 28, 60\}$ GHz \\ 
		Velocity MSE Analysis & $n \in \{200, 800, 3200\}$ \\ 
		\bottomrule
	\end{tabularx}
\end{table}
\section{Numerical Results}\label{sec.VI}
This subsection investigates the tradeoffs between the sensing error $e_{th}$ and the communication code rate $\frac{b}{n}$ under various system configurations. The average SNR at the transmitter side  is defined as
	$\mathrm{SNR} = \frac{\bar{p}}{\sigma_{n}^2}$,
where the SNR at the receiver side can be obtained by considering the channel variance $\sigma_{H}^2\mathrm{SNR}$.
\subsubsection{Illustration of Tradeoff}
Fig.~\ref{fig:coderate_SNR_fixed_sensing} presents the achievability and converse bounds of the code rate under fixed sensing error $e_{th}\in \{10^{-3},10^{-2},5\times 10^{-3},2\times 10^{-2},5\times 10^{-2}\}$, $k=16$ active users, $n=1000$ channel uses, and $m=10$ receiving antenna over a range of SNR values. As expected, the converse bound is constrained by the Shannon capacity. With a smaller sensing precision threshold $e_{th}$, the converse bound approaches the Shannon capacity at a lower SNR. For the achievability bound, all curves converge to an asymptotic line under a fixed number of active users, which is consistent with the behavior predicted by \eqref{eq:perm_cor_achi}. Under fixed $k$, the permissible correlation is given by $\widehat{\rho}_{\mathrm{chi}}=\frac{1}{k-1} \left( 1 - \frac{e_{\min}}{e_{th}} \right)$, which converges to $\widehat{\rho}_{\mathrm{chi}}=\frac{1}{k-1}$ as the SNR increases and $e_{\min}\rightarrow 0$. The region between the converse bound and the achievability bound represents the available space for further optimization.

In Fig.~\ref{fig:achi_conv_wall}, the communication--sensing tradeoff under different blocklengths is illustrated to reveal the impact of blocklength on both the proposed achievability bound and the converse bound. The results are obtained for $k=16$ active users, $m=10$ receiving antennas, and $\mathrm{SNR}=10\,\mathrm{dB}$. For both bounds, a \emph{silent region} is observed, indicating that the sensing requirement is so stringent that reliable communication cannot be supported under the given energy constraint, regardless of the blocklength. Increasing the blocklength only shrinks this silent region. However, it does not eliminate it due to the inherent strictness imposed by the sensing requirement. As the sensing threshold $e_{th}$ decreases, the achievability bound gradually converges to the asymptotic surface, and the converse-bound surface remains fundamentally limited by Shannon capacity constraints.

\subsubsection{Illustration of Practical Sensing Examples}
To demonstrate the universality of the lower bound in \eqref{eq:MSE_param} for sensing parameters derived from channel estimation, Fig.~\ref{fig:practical_parameter_estimation} presents the MSE performance of AoA estimation $\theta_i$, range estimation $r$, and velocity estimation $v$. The corresponding Jacobian derivations have been detailed in Sec.~\ref{sec.3GPP}, and the simulation parameters are summarized in Tab.~\ref{tab:sim_params}. As these parameters exhibit different sensitivities to the available degrees of freedom across distinct subspaces, multiple system configurations are considered. Specifically, Fig.~\ref{fig:practical_parameter_estimation} evaluates the MSE lower bounds of AoA, range, and velocity estimation under varying numbers of receiving antennas, frequency bands, and blocklengths, respectively. The results validate the effectiveness of using channel estimation as a unified, high-level sensing metric.

\section{Conclusion}\label{sec.VII}
This work studied uplink dual-functional ISAC multiple access under finite blocklength constraints and characterized its fundamental communication--sensing tradeoffs. A geometric interpretation of the sensing error revealed that codeword correlation plays a decisive role in limiting sensing accuracy and achievable rate. Achievability and converse bounds were derived to quantify this tradeoff, while a universal CRB connected channel state sensing to practical sensing parameters. The results provide theoretical insights for designing ISAC multiple-access systems with short-packet or low-latency requirements.

\section*{Acknowledgment}
\smaller
This work of J. Dang, Z.-C. Zhang is supported by the Fundamental Research Funds for the Central Universities (2242022k60001), Basic Research Program of Jiangsu (No. BK20252003, the Key Laboratory of Intelligent Support Technology for Complex Environments, Ministry of Education, Nanjing University of Information Science and Technology (No. B2202402). The work of K. Meng is supported in part by UKRI under Grant EP/Y02785X/1.

\balance

	%
	%
	%
	%
	%

	\vfill
	

\begin{thebibliography}{1}
		\smaller
		\bibliographystyle{IEEEtran}
		\bibitem{ISAC1}
		F. Liu, {\em et al.}, ``Integrated sensing and communications: Toward dual-functional wireless networks for 6G and beyond,'' {\em IEEE J. Sel. Areas Commun.
		}, vol. 40, no. 6, pp. 1728-1767, Jun. 2022.
		\bibitem{ISAC3}
		J. A. Zhang, , {\em et al.}, ``An overview of signal processing techniques for joint communication and radar sensing,'' {\em IEEE J. Sel. Topics Signal Process.}, vol. 15, no. 6, pp. 1295-1315, Nov. 2021.
		\bibitem{ISAC_Mulitple_Access}
		Y. Liu, , {\em et al.}, ``Next-generation multiple access for integrated sensing and communications,'' {\em Proc. IEEE}, vol. 112, no. 9, pp. 1467-1496, Sept. 2024.
		\bibitem{TMDA1}
		E. Shayo, {\em et al.}, ``A survey on time division multiple access scheduling algorithms for industrial networks,'' {\em Social Netw. Appl. Sci.}, vol. 2, no. 12, p. 2140, Dec. 2020.

		\bibitem{NOMA-ISAC}
		L. Sun, {\em et al.}, ``On the study of non-orthogonal multiple access (NOMA)-assisted integrated sensing and communication (ISAC),'' {\em IEEE Trans. Commun.}, vol. 72, no. 11, pp. 7278-7293, Nov. 2024.
		\bibitem{ISAC-fundamentals}
		A. Liu, {\em et al.}, ``A survey on fundamental limits of integrated sensing and communication,'' {\em IEEE Commun. Surv. Tutorials}, vol. 24, no. 2, pp. 994-1034, Secondquarter 2022.
		
		
		\bibitem{Polyanskiy}
		Y. Polyanskiy, {\em et al.}, ``Channel coding rate in the finite blocklength regime,'' {\em IEEE Trans. Inf. Theory}, vol. 56, no. 5, pp. 2307–2359, May 2010.
		\bibitem{FBL}
		P. Mary, {\em et al.}, ``Finite blocklength information theory: What is the practical impact on wireless communications?,'' {\em Proc. Proc. IEEE Global Commun. Conf. (GLOBECOM) Workshop}, Washington, DC, USA, 2016, pp. 1-6.
		
		
		\bibitem{FBL-ISAC2}
		H. Nikbakht, {\em et al.}, ``Integrated sensing and communication in the finite blocklength regime,'' in {\em Proc. IEEE Int. Symp. Inf. Theory (ISIT)}, Athens, Greece, 2024, pp. 2790-2795.
		\bibitem{FBL-ISAC1}
		X. Shen, {\em et al.}, ``On the performance tradeoff of an ISAC system with finite blocklength,'' in {\em Proc. IEEE Int. Conf. Commun. (ICC)}, Rome, Italy, 2023, pp. 4628-4633.
		\bibitem{FBL-ISAC3}
		X. Shen, {\em et al.}, ``Fundamental tradeoff of bistatic ISAC under gaussian fading channels at finite blocklength,'' {\em IEEE Trans. Inf. Theory}, {\em Early Access}, \url{doi: 10.1109/TIT.2025.3623189}.
		
	
		
		\bibitem{Theorem3}
		Z. Zhang, {\em et al.}, ``Finite-blocklength fluid antenna systems,'' 	\url{arXiv:2509.15643}, 2025.
		
		\bibitem{GCT}
		R. A. Horn and C. R. Johnson, {\em Matrix Analysis}, 2nd ed. Cambridge, U.K., Cambridge Univ. Press, 2013.

		\bibitem{Funda_wireless}
		D. Tse and P. Viswanath, {\em Fundamentals of Wireless Communication}. Cambridge, U.K.: Cambridge Univ. Press, 2005.
		\bibitem{Concavity}
		E. Telatar, ``Capacity of multi-antenna gaussian channels,'' {\em Eur. Trans. Telecommun.}, vol. 10, no. 6, pp. 585-595, Nov.-Dec. 1999.
		\bibitem{3gpp_38901}
		3GPP, {\em Study on channel model for frequencies from 0.5 to 100 GHz}, 3rd Generation Partnership Project (3GPP), Tech. Rep. 38.901 V17.0.0, Mar. 2022.
		\bibitem{channel_modeling}
		H. Jiang, \emph{et al.}, ``Dynamic channel modeling of fluid antenna systems in UAV communications,'' \emph{IEEE Wireless Commun. Lett.}, vol. 14, no. 10, pp. 3169-3173, Oct. 2025.
	\end{thebibliography}
\end{document}